\documentstyle[prl,twocolumn,aps]{revtex}
\begin{document}
\title
{Aharonov-Bohm oscillations and resonant tunneling in
strongly correlated quantum dots}
\author{C. Bruder$^1$, Rosario Fazio$^{1,2}$, and
Herbert Schoeller$^{1,3}$}
\address{
$^1$ Institut f\"ur Theoretische Festk\"orperphysik, Universit\"at
Karlsruhe, 76128 Karlsruhe, Germany\\
$^2$ Istituto di Fisica, Facolt\`a di Ingegneria, viale A. Doria 6,
95129 Catania, Italy\\
$^3$ Department of Physics, Simon Fraser University, Burnaby, B.C.,
V5A 1S6, Canada}
\date{\today}
\maketitle
\begin{abstract}
We investigate Aharonov-Bohm oscillations of the current
through a strongly correlated quantum dot embedded in an
arbitrary scattering geometry. Resonant-tunneling
processes lead to a flux-dependent renormalization of
the dot level. As a consequence we obtain a fine structure
of the current oscillations which is controlled by quantum
fluctuations. Strong Coulomb repulsion leads to a continuous
bias voltage dependent phase shift and, in the nonlinear response regime,
destroys the symmetry of the differential conductance under a sign
change of the external flux.
\end{abstract}

Phase-sensitive transport properties of interacting mesoscopic systems are
important for several reasons. The small size of the samples gives rise
to capacitances of order $10^{-15} F$ which induce Coulomb blockade effects
\cite{Ave-Lik} and demand the necessity to generalize the Landauer-B\"uttiker
formalism \cite{Lan} to systems with strong interactions. Furthermore, the
investigation of Aharonov-Bohm oscillations through quantum dots
with strong Coulomb repulsion might give further experimental evidence for
resonant tunneling and Kondo phenomena in nonequilibrium systems
\cite{Ng-Lee,Koe-Sch-Sch,Ral-Buh}.

Interference effects in the Coulomb blockade regime have been measured by
Yacoby {\it et al.} \cite{Yac-etal} by studying a quantum dot embedded in
an Aharonov-Bohm ring. This experiment demonstrates that phase-coherent
transport through quantum dots is possible in realistic experiments and
is not destroyed by inelastic interactions. Recent theoretical work on
Aharonov-Bohm oscillations in a mesoscopic ring with a quantum dot
\cite{Yey-Bue,Hac-Wei} uses a noninteracting model. Using the symmetry of
the current under sign change of the external flux in the linear
response regime \cite{Gef-Imr-Azb,Bue-ibm}
it was shown in Ref.~\cite{Yey-Bue} that the phase of the Aharonov-Bohm
oscillations can only take two possible values as function of the gate
voltage on the dot. However, the experiment of Yacoby {\it et al.} is
performed in the Coulomb blockade regime where interaction effects are
important. In this letter we will take such correlations into account by
setting up a complete and general theory for interference phenomena in
strongly interacting quantum dots embedded in an arbitrary noninteracting
multi-probe and multi-channel scattering structure. As a consequence we will
show that the symmetry under sign change of the external flux
is broken in the nonlinear response regime and
that the phase can change continuously as a function of the bias voltage.
Furthermore, we will analyze in detail the current oscillations as a function
of the gate voltage caused by the flux-modulated renormalization
of the local energy level of the dot.

To have a specific example, we will study the system shown in Fig.~\ref{fig1}
although our formalism is valid for an arbitrary scattering geometry. For
simplicity we start with the case of one-channel leads. The system without
the quantum dot is described by scattering waves with zero boundary
conditions at the tunneling barriers of the dot. Thus, in energy
representation, this part of the Hamiltonian is given by
$H_S = \sum_{\alpha\sigma} \int d\epsilon\,\epsilon\,
a^\dagger_{\alpha\sigma}(\epsilon)\,a_{\alpha\sigma}(\epsilon)\; ,$
where $a^\dagger_{\alpha\sigma}(\epsilon)$ creates an incoming
scattering wave in probe $\alpha$ with spin $\sigma$
and total energy $\epsilon$. The isolated dot is described
by
$H_D = \sum_\sigma\,\epsilon_\sigma\,d^\dagger_\sigma d_\sigma
\,+\,U\sum_{\sigma<\sigma^\prime}\,n_\sigma n_{\sigma^\prime}$
with single particle energies $\epsilon_\sigma$ and on-site repulsion $U$.
The position of the dot levels are controlled by an external gate voltage
and $U\sim 1-5$K corresponds to the charging energy \cite{Bru-Sch}.

The tunneling of the electrons into or out of the dot is described
by
\begin{equation}\label{4}
H_T = \sum_{\alpha\sigma}\int d\epsilon\,\left\{ t_\alpha
(\epsilon)\,a^\dagger_{\alpha\sigma}(\epsilon)\,d_\sigma
\, + \, h.c.\right\}\; .
\end{equation}
Here, $t_\alpha (\epsilon) = \sum_{i=L/R} t_i\,\langle\alpha
\epsilon|x_i\rangle$ are the tunneling matrix elements in energy
representation, where $t_i$ are real quantities and
$\langle x|\alpha\epsilon\rangle$ is the spin-independent scattering
wave from reservoir $\alpha$ with energy $\epsilon$ at
position $x$. By $x_i$ , $i=L,R$, we denote an arbitrary point
in the one-dimensional left or right lead which is connected
to the dot \cite{com1}. Due to zero boundary conditions
we have $\langle x_i|\alpha\epsilon\rangle=\rho(\epsilon)^{1/2}
A^i_\alpha(\epsilon) \sin(k(\epsilon) x_i)$ with the one-dimensional
density of states $\rho(\epsilon)=1/(\pi\hbar v(\epsilon))$ and
energy $\epsilon=\hbar^2 k(\epsilon)^2/(2m)={1\over 2} m
v(\epsilon)^2$. The coefficients $A^i_\alpha$ depend on the detailed
scattering problem under consideration.
We have chosen the tunneling matrix elements $t_i$ as real
parameters which means that we shift the complete flux
dependence to the scattering Hamiltonian $H_S$ via a
standard gauge transformation.

Following B\"uttiker \cite{Bue}, we will use the following
representation of the current operator in probe $\alpha$
\begin{equation}\label{5}
\hat{I}_\alpha = {e\over h}\int d\epsilon d\epsilon^\prime
\sum_\sigma\left[a^\dagger_{\alpha\sigma}(\epsilon)
a_{\alpha\sigma}(\epsilon^\prime) - b^\dagger_{\alpha\sigma}
(\epsilon)b_{\alpha\sigma}(\epsilon^\prime)\right]\; ,
\end{equation}
where $b_{\alpha\sigma}(\epsilon)=\sum_\beta
s_{\alpha\beta}(\epsilon) a_{\beta\sigma}(\epsilon)$ annihilates an
outgoing carrier in probe $\alpha$ and $s$
is the scattering matrix of the system without the dot.
To calculate the average current
$I_\alpha = \langle \hat{I}_\alpha \rangle$ in the
stationary limit we need the stationary real-time Green's function
$G^<(E)=\int dt e^{iEt} G^<(t)$ in Fourier space
of two scattering field operators: $G^<_{\alpha\sigma,
\alpha^\prime\sigma^\prime}(\epsilon,\epsilon^\prime;t) =
i\langle a^\dagger_{\alpha\sigma}(\epsilon,t)\,a_{\alpha^\prime
\sigma^\prime}(\epsilon^\prime)\rangle$. Using the matrix notation
\begin{math}
\hat{G}=\left(\begin{array}{ccc} G^R & G^< \\ 0 & G^A
\end{array}\right)
\end{math},
where $G^R$ and $G^A$ are the retarded and advanced Green's
functions, and applying the Keldysh technique \cite{Kel},
we can express the Green's function $\hat{G}_{\alpha\sigma,
\alpha^\prime\sigma}$ exactly by
the local Green's function $\hat{G}_{\sigma}$ of the dot,
$\hat{G}_{\alpha\sigma,\alpha^\prime\sigma}(\epsilon,
\epsilon^\prime;E) =
\hat{g}_\alpha (\epsilon;E)\delta_{\alpha,\alpha^\prime}
\delta(\epsilon-\epsilon^\prime)
+t_\alpha (\epsilon)
t_{\alpha^\prime}(\epsilon^\prime)^* \hat{g}_\alpha (\epsilon;E)
\hat{G}_{\sigma}(E)\hat{g}_{\alpha^\prime}(\epsilon^\prime;E)\; ,$
where we have already used spin conservation. The Green's
functions $\hat{g}_\alpha$ correspond to the Hamiltonian
$H_S$ and are given by $g_\alpha^{R/A}(\epsilon;E)=
(E-\epsilon\pm i0^+)^{-1}$ and $g_\alpha^< (\epsilon;E)
=2\pi i f_\alpha (E)\delta(E-\epsilon)$ where $f_\alpha$
is the Fermi distribution function of reservoir $\alpha$.
Using this result in calculating the average current,
inserting the form of the tunneling matrix elements and
performing the energy integrations \cite{com2}, we obtain
\begin{eqnarray}\nonumber
I_\alpha &=& I^{(0)}_\alpha + {e\over h} Re
\sum_\sigma\sum_{\beta\gamma}\int dE \\ &&\times
s^\dagger_{\alpha\beta}\,s_{\alpha\gamma}\,A_{\gamma\beta}
\,({i\over 2} G^<_{\sigma} + if_\beta G^R_\sigma)\; ,
\label{7}
\end{eqnarray}
where $I^{(0)}_\alpha$ is the current without the dot (given
by the usual Landauer-B\"uttiker formula) and
$A_{\alpha\alpha^\prime}=\sum_{ij}(\Gamma_i\Gamma_j)^{1\over 2}
{A^i_\alpha}^*A^j_{\alpha^\prime}$ with $\Gamma_i(\epsilon)=
2\pi\rho (\epsilon) t_i^2$.

Eq.~(\ref{7}) is the first central result of this
paper. It relates the current in probe $\alpha$ exactly
to the local Green's functions of the dot and the scattering
properties of the noninteracting medium surrounding the dot.
This formula is a natural generalization of the Landauer-B\"uttiker
formula to an interacting quantum dot. Furthermore it
generalizes current formulas through
quantum dots connected to two leads without any
possibility of a direct transition between the probes
\cite{Mei-Win}. Here we are able to account for such transitions
opening the possibility to study interference phenomena in
the presence of locally interacting subsystems. The
generalization of Eq.~(\ref{7}) to multi-channel leads is
straightforward. Again following Ref.~\cite{Bue} the field
operators $a_{\alpha\sigma}$ in Eq.~(\ref{5}) have to be
treated as vectors with a channel index $n$. Equivalently,
the matrix elements $s_{\alpha\beta}$ and $A_{\alpha\beta}$
have to be treated like $Z_\alpha \times Z_\beta$ matrices where
$Z_\alpha$ is the number of transverse channels in lead $\alpha$.
The final formula for the current is then exactly like
Eq.~(\ref{7}) except that we have to take the trace of the
matrix multiplication $s^\dagger_{\alpha\beta}s_{\alpha\gamma}
A_{\gamma\beta}$.

The scattering matrices in Eq.~(\ref{7}) can be found by
straightforward quantum-mechanical considerations depending
on the specific geometry. For the Green's functions of the dot
we will use a real-time technique developed in
Ref.~\cite{Sch-Sch} which has been applied to a quantum dot
in Ref.~\cite{Koe-Sch-Sch}.
For a degenerate dot level (i.e., $\epsilon_\sigma=
\epsilon_d$ independent of spin) and in the $U=\infty$ limit, one obtains
$G^{</>}_\sigma (E) = 2\pi i\,\gamma^{\pm}(E)
|E-\epsilon_d-\sigma (E)|^{-2}$. Here,
\begin{equation}\label{8}
\sigma(E) = \int dE^\prime {M\gamma^+(E^\prime)+
\gamma^-(E^\prime) \over E-E^\prime + i0^+}
\end{equation}
has the form of a self-energy which describes the
renormalization and broadening of the dot level
$\epsilon_d$, $\gamma^{\pm}(E)=\sum_\alpha
|t_\alpha (E)|^2 f_\alpha^\pm (E)=1/(4\pi)\sum_\alpha
A_{\alpha\alpha}(E) f_\alpha^\pm(E)$ is the
classical rate for a particle tunneling in or out
of the dot, and $f_\alpha^+ = f_\alpha$ while
$f_\alpha^- = 1-f_\alpha$. The retarded Green's
function follows from $Im\ G^R = 1/(2i)
(G^> - G^<)$ and the real part is obtained from the
Kramers-Kronig relation.

The explicit result for the Green's functions together with the expression
(\ref{7}) for the current constitutes a complete theory of interference
effects in mesoscopic scattering geometries with an interacting part given by
a quantum dot with one degenerate level. Our result satisfies current
conservation $\sum_\alpha I_\alpha = 0$, and all currents vanish in
equilibrium. Furthermore, for the special case $M=1$ where the Coulomb
interaction does not play any role, our result is exact and can be shown to
agree with the Landauer-B\"uttiker formalism.

The real part of the self-energy $\sigma$ describes the renormalization of
the dot level. If we neglect the energy dependence of $A_{\alpha\alpha}$ at
the Fermi level, we obtain from Eq.~(\ref{8}) for a two-terminal system
\begin{eqnarray}\nonumber
Re\ \sigma &=& Re\ \sigma_1 + {M-1\over 8\pi}
\left [ (A_{11}+A_{22})(\chi_1+\chi_2)+\right.\\
&&\hspace{2cm}+\left.(A_{11}-A_{22})
(\chi_1-\chi_2)\right]\; ,
\label{9}
\end{eqnarray}
where $\sigma_1$ is the self-energy for $M=1$ and
$\chi_\alpha (E)=Re\int dE^\prime f_\alpha (E^\prime)/
(E-E^\prime+i0^+)$. Using a Lorentzian cutoff at
$D$ (which will be of the order of the Coulomb repulsion
$U$), we obtain $\chi_\alpha (E) = \psi ({1\over 2}+{D\over 2\pi T})
-Re\ \psi ({1\over 2} + i{E-\mu_\alpha \over 2\pi T})$
where $\psi$ is the digamma function and $\mu_\alpha$
the chemical potential of reservoir $\alpha$.
$\sigma_1$ is always a symmetric function of the
external flux $\Phi$. Furthermore, for a spatially symmetric
situation as in Fig.~\ref{fig1}, $A_{11}\pm A_{22}$
is an even (odd) function of the phase $\varphi=2\pi\Phi/\Phi_0$
($\Phi_0$ being the flux quantum). Due to $Re\ \sigma_1$
the level position of the dot will
oscillate with $\varphi$ with an amplitude of the
order of $\Gamma$ and phase 0 or $\pi$. For $M>1$,
there can be logarithmic corrections in temperature and
bias voltage for the amplitude {\it and}
phase of this oscillation due to the $\chi_\alpha$
functions. The latter terms usually lead to Kondo-like
correlations \cite{Ng-Lee,Koe-Sch-Sch}.

To exhibit the consequences of the oscillation of the renormalized dot level
we will now apply our results to the specific scattering geometry of
Fig.~\ref{fig1} which corresponds to the experimental setup of
Ref.~\cite{Yac-etal}. For simplicity we assume a one-dimensional structure
and we use the same scattering matrices $s^{i,o}$ for the incoming and
outgoing junctions as in Ref.~\cite{Gef-Imr-Azb}.
The scattering matrix of the upper arm (including the
flux and the phases accumulated by free motion) is written in the form
\begin{math}
s^T=p\left(\begin{array}{ccc} r & t e^{-i\varphi} \\
t e^{i\varphi} & r
\end{array}\right)
\end{math},
where $p=e^{ikl}$ is the phase acquired by free motion through the upper arm.
Furthermore, we take the length of the leads connected to the quantum dot
as $l_L=l_R={1\over 2}l$ and we assume a symmetric quantum dot with
$\Gamma_L=\Gamma_R=\Gamma$.

We will look explicitly at two cases, viz., perfect transmission
through the upper arm given by $r=0$, $t=1$, or weak
transmission described by $r=-1$, $t=i|t|$. In the first
case we obtain after a straightforward calculation
$s_{11}=s_{22}={1\over 2} p(p-1)$, $s_{12}(\varphi)
=s_{21}(-\varphi)={1\over 2}p(p+1)e^{i\varphi}$,
$A^L_1=A^R_2={1\over 2}i\sqrt{p}(2-p)$ and $A^L_1(\varphi)=
A^R_2(-\varphi)=-{1\over 2} ip\sqrt{p}e^{i\varphi}$.
In the second case one obtains
$s_{11}=s_{22}=-p$, $s_{12}(\varphi)=s_{21}(-\varphi)
={1\over 2}ip|t|e^{i\varphi}$, $A^L_1=A^R_2=i\sqrt{p}$
and $A^L_1(\varphi)=A^R_2(-\varphi)={1\over 2}{p\sqrt{p}
\over 1+p}|t|e^{i\varphi}$.
The phase $p$ cannot be determined and will have some specific value in the
experiment. We assume here always to be
in the quantum region, i.e., the lengths associated with temperature, bias
voltage, and $\Gamma$ should exceed the system length, so that we can neglect
the energy dependence of $p$. In the perfect transmission case we choose
$p=i$ and get
$Re\ \sigma_1 = -{1\over 4}\Gamma
(3+\cos{\varphi})$, $A_{11}+A_{22}=\Gamma (3+\cos{\varphi})$
and $A_{11}-A_{22}=-2\Gamma \sin{\varphi}$.
For weak transmission we take $p=1$ and obtain
$Re\ \sigma_1 = -{1\over 8}|t|\Gamma
\cos{\varphi}$, $A_{11}+A_{22}=2\Gamma$ and $A_{11}-A_{22}=-|t|
\Gamma\sin{\varphi}$.

In Fig. \ref{fig2} we show the linear conductance for perfect transmission
and $T=\Gamma=0.01$ (in units of the cutoff $D$) for various positions of the
dot level. We have assumed $M=2$, i.e., the interacting case. The linear
conductance is symmetric under a sign change of the flux since the
$\sin{\varphi}$ term is absent in Eq.~(\ref{9}) for $\chi_1=\chi_2$.
If the position of the dot level
is below $\epsilon_d \approx -0.02$, the current has a maximum around
$\varphi=0$. For higher values of $\epsilon_d$, the current has a minimum
at $\varphi=0$. Although this looks similar to the abrupt phase change of
$\pi$ described in Ref. \cite{Yac-etal}, Fig. \ref{fig2} shows that the
response of the system cannot be described by the concept of a
``phase shift''. What happens instead is that the current (as a function
of $\varphi$) changes its functional form. The scale of the transition is
independent of temperature and is given by the intrinsic parameter
$\Gamma$. For higher temperatures we
find a smaller amplitude of the current oscillations but the qualitative
picture remains. For weak transmission, the results are similar
but the scale is given by $t\Gamma$.

We will now turn to the nonlinear conductance. Figure \ref{fig3} depicts the
differential conductance for weak transmission as a function
of $\varphi$ for different voltages and level positions ($T=10\Gamma=0.1$).
For $V=0$, the differential conductance is symmetric around $\varphi=0$,
in the nonlinear response case, this symmetry is absent due to the
last term in Eq.~(\ref{9}). The
behavior of the differential conductance at the origin changes from a
minimum to a maximum as a function of the level position $\epsilon_d$,
this time rather abruptly (energy scale $t\Gamma$). The asymmetry of the
conductance curves for finite voltages is a genuine interaction effect;
it disappears for $M=1$. Furthermore we observe a continuous phase
shift of the current oscillations as function of the bias voltage which
again is absent for $M=1$. It is determined by the last term
in Eq.~(\ref{9}) as well as by $\sin{\varphi}$ terms
occuring explicitly in the current formula (\ref{7}) via the
$A_{\gamma\beta}$ matrices. Note that the temperature is one order of
magnitude larger than $\Gamma$ in this figure, i.e., an interference
experiment of this type might yield information about correlation
effects at temperatures which are accessible in experiments \cite{Fox}.

Finally we want to comment on the influence of
interactions on the relative phase of the
Aharonov-Bohm oscillations at successive peaks in
the linear conductance as function of the gate
voltage. In a noninteracting model two adjacent
peaks correspond to transport through two different
energy levels of the dot which have different
parity. Thus the relative sign of $t_L$ and
$t_R$ would change from one level to the next
and consequently one expects a phase shift of
$\pi$. However, in the experiment of Yacoby
{\it et al.} no phase shift was measured. In addition
to the discussion of Ref.~\cite{Yey-Bue},
a strong Coulomb repulsion on the quantum dot could be
an explanation for this observation. If there are $N$ states
on the dot which lie close together in energy
but with the same parity in longitudinal direction
(e.g. spin degenerate states or states differing
in the transverse channel number),
there would be $N$ adjacent Coulomb
peaks with the same phase of the Aharonov-Bohm
oscillations. The distance of these Coulomb peaks
is given by the charging energy $U$ whereas in
the noninteracting case all these peaks would
fall together into one single peak. Therefore
we conclude that in the presence of interactions
the parity of the energy levels contributing to
transport at adjacent Coulomb peaks can be the
same which provides an explanation for Yacoby's experiment.

In conclusion, we have presented a complete theory for interference
phenomena in strongly correlated quantum dots embedded in a scattering
geometry. On one hand, we have
found that the functional form of ${dI \over dV}(\varphi)$ is changing with
the gate voltage on the scale of temperature-independent intrinsic parameters.
In linear response this change cannot be interpreted as a phase shift. On
the other hand, we have shown that in the nonlinear response regime,
correlation effects break the symmetry
under sign change of the external flux and lead to a real continuous phase
shift as a function of the bias voltage.

\acknowledgements
It is a pleasure to thank G. Hackenbroich for initiating this work and for
stimulating discussions. We would also like to acknowledge helpful discussions
with M. B\"uttiker, Y. Imry, J. K\"onig, and A. Yacoby.
This work was supported by the Deutsche Forschungsgemeinschaft through
Sonderforschungsbereich 195 and by the Swiss National Science Foundation
(H.S.).

\begin{figure}
\vspace{0.3cm}
\caption{Geometry of the model system studied here. Leads 1 and 2 connect
to the left and right reservoir (shaded). The ring is connected to a
quantum dot via high tunneling barriers. $\Phi$ is the flux penetrating
the ring.}
\label{fig1}
\end{figure}

\begin{figure}
\vspace{0.3cm}
\vspace{0.3cm}
\caption{Linear conductance (in units of $e^2/h$) as a function of
magnetic flux for various positions of
the level in the dot ($T=\Gamma=0.01$, $M=2$). We have assumed perfect
transmission through the upper arm of the system as in the experiment by
Yacoby {\it et al.} \protect\cite{Yac-etal}.
}
\label{fig2}
\end{figure}

\begin{figure}
\vspace{0.3cm}
\caption{Differential conductance (in units of $e^2/h$) for the case of
weak transmission ($t=0.1$) as a function of magnetic flux for various
voltages and positions of the level in the dot ($T=10\Gamma=0.1$, $M=2$).
Note that the linear conductance ($V=0$) is symmetric around $\varphi=0$,
whereas there is no such symmetry in the nonlinear response case.
}
\label{fig3}
\end{figure}


\begin{references}
\bibitem{com1}
The position of the tunnel junction is chosen as
reference point $x=0$. In the final equations we
will always set $x_L=x_R$ and neglect oscillating
terms in $x_{L/R}$.
\bibitem{com2}
The integrations over $\epsilon$ and $\epsilon^\prime$
in (\protect\ref{5}) can be performed analytically since
one can prove that $A^i_\alpha (\epsilon)$ and
$\sum_\beta s_{\alpha\beta}(\epsilon)A^i_\beta
(\epsilon)^*$ are analytic functions in the upper
half of the complex plane.
\end{references}
\end{document}